\documentstyle[12pt]{article}
\def\lsim{\:\raisebox{-0.5ex}{$\stackrel{\textstyle<}{\sim}$}\:}
\def\gsim{\:\raisebox{-0.5ex}{$\stackrel{\textstyle>}{\sim}$}\:}
\def\mathrm#1{\hbox{#1}}
\newcommand {\ignore}[1]{}

\newcommand{\noi}{\noindent}
\newcommand{\bc}{\begin{center}}
\newcommand{\ec}{\end{center}}
\newcommand{\epm}{e^+e^-}
\def\ifmath#1{\relax\ifmmode #1\else $#1$\fi}
\def\3quarter{{\textstyle{3 \over 4}}}

\def\vs{\vskip}

\def\ra{\rightarrow}

\def\lf{\leaders\hbox to 1em{\hss.\hss}\hfill}

\def\21{$SU(2) \ot U(1)$}
\def\321{$SU(3) \ot SU(2) \ot U(1)$}

\def\nt{\hbox{$\nu_\tau$ }}

\def\gau{\hbox{gauge }}

\def\neu{\hbox{neutrino }}

\def\eq#1{{eq. (\ref{#1})}}

\def\VEV#1{\left\langle #1\right\rangle}

\def\lsim{\raise0.3ex\hbox{$\;<$\kern-0.75em\raise-1.1ex\hbox{$\sim\;$}}}
\def\gsim{\raise0.3ex\hbox{$\;>$\kern-0.75em\raise-1.1ex\hbox{$\sim\;$}}}

\def\bel{\begin{letter}}
\def\eel{\end{letter}}
\def\beq{\begin{equation}}
\def\eeq{\end{equation}}
\def\bef{\begin{figure}}
\def\eef{\end{figure}}
\def\bet{\begin{table}}
\def\eet{\end{table}}
\def\bea{\begin{eqnarray}}
\def\ba{\begin{array}}
\def\ea{\end{array}}
\def\bi{\begin{itemize}}
\def\ei{\end{itemize}}
\def\ben{\begin{enumerate}}
\def\een{\end{enumerate}}
\def\ra{\rightarrow}
\def\ot{\otimes}

\def\eea{\end{eqnarray}}
\def\ib#1#2#3{           {\it ibid. }{\bf #1} (19#2) #3}

\def\nps#1#2#3{          {\it Nucl. Phys. B (Proc. Suppl.) }
                         {\bf #1} (19#2) #3}
\def\np#1#2#3{           {\it Nucl. Phys. }{\bf #1} (19#2) #3}
\def\pl#1#2#3{           {\it Phys. Lett. }{\bf #1} (19#2) #3}
\def\pr#1#2#3{           {\it Phys. Rev. }{\bf #1} (19#2) #3}
\def\prep#1#2#3{         {\it Phys. Rep. }{\bf #1} (19#2) #3}
\def\prl#1#2#3{          {\it Phys. Rev. Lett. }{\bf #1} (19#2) #3}

\def\zp#1#2#3{           {\it Zeit. fur Physik }{\bf #1} (19#2) #3}

\def\n.c.#1#2#3{         {\it Nuovo Cim. }{\bf #1} (19#2) #3}
\def\r.n.c.#1#2#3{       {\it Riv. del Nuovo Cim. }{\bf #1} (19#2) #3}

\def\mpl#1#2#3{          {\it Mod. Phys. Lett. }{\bf #1} (19#2) #3}

\relax
\begin{document}
\begin{titlepage}
\rightline{FTUV/95-02}
\rightline{IFIC/95-02}
\rightline{hep-ph yymmdd}
\rightline{January 1995}
\noindent
\begin{center}
{\Large \bf SINGLE PHOTON DECAYS OF THE $Z^0$ AND SUSY WITH
SPONTANEOUSLY BROKEN R-PARITY}\\
\vskip .4cm
{\large \bf J. C. Rom\~ao}
\footnote{e-mail fromao@alfa.ist.utl.pt}\\
Instituto Superior T\'ecnico, Dept. de F\'{\i}sica \\
Av. Rovisco Pais 1, 1096 Lisboa Codex, PORTUGAL\\
{\large \bf J. Rosiek}
\footnote{Bitnet ROSIEK@EVALVX and ROSIEK@fuw.edu.pl}\\
and\\
\vs .1cm
{\large \bf J. W. F. Valle}
\footnote{Bitnet VALLE@EVALUN11 - Decnet 16444::VALLE}\\
Instituto de F\'{\i}sica Corpuscular - IFIC/C.S.I.C.\\
Dept. de F\'{\i}sica Te\`orica, Universitat de Val\`encia\\
46100 Burjassot, Val\`encia, SPAIN\\
\vs .1cm
\end{center}
\vskip 1.0cm
\begin{abstract}
\baselineskip=14pt
{Spontaneous violation of R parity can induce rare
single photon decays of the $Z^0$ involving the emission
of (nearly) massless pseudoscalar Goldstone bosons,
majorons, as well as massive CP even or CP odd spin
zero bosons that arise in the electroweak breaking
sector of these models. We show that the majoron
emitting decays can have a sizeable
branching ratio of $10^{-5}$ or so, without
conflicting any experimental observation from
neutrino physics or particle searches.
These decays may lead to interesting structures for
the single photon spectrum involving either mono
chromatic photons as well as continuous spectra
that grow with energy. They would easily account
for an excess of single photon events at high
energies recently hinted at by the OPAL
collaboration.
}
\end{abstract}

\end{titlepage}

\section{Introduction}

One of the classic missing energy experiments in $e^{+} e^{-}$
annihilation is the neutrino counting with single photon events
\cite{opal}. Of course, such events are expected to occur
via initial state bremsstrahlung with the $Z$ decaying to a
$\nu \bar{\nu}$ pair. While the accurate measurements of the $Z$
lineshape at LEP have now achieved a much better accuracy for counting
the number of light neutrino species, the single photon energy
spectrum still remains interesting. Recently the OPAL collaboration
has published a high statistics single photon spectrum that shows
some excess of high energy photons above the expectations from
initial state radiation.

There has been a lot of interest on the phenomenology of
supersymmetric models. The most well studied scheme has
so far been the so-called minimal supersymmetric standard
model \cite{mssm}, which assumes the conservation of a
discrete R parity symetry \cite{weinb}. In this model,
in addition to the single photon events expected from
initial state radiation (as in the standard model)
one expects some excess of high energy photons
from the production of two neutralinos,
followed by the radiative decay of the heaviest to
the lightest (LSP). In this case the missing momentum
is carried by two LSPs and requires $m_Z \geq 2 m_{LSP}$.

In this paper we note some interesting features related to single
photon events which are expected in a class of SUSY models with
spontaneous violation of R parity \cite{MASI_pot3,MASI2}. These
models are characterized by the existence  of a massless (or nearly
so) pseudoscalar Goldstone boson, called majoron \cite{CMP}, that
follows from the spontaneous nature of the underlying lepton number
violation.

These models predict the existence of new decay modes for this
$Z$-boson involving single majoron emission
$$
Z^0 \rightarrow J \gamma
$$
\noindent
where J denotes the majoron. Since the pseudoscalar boson
is massless (or nearly so) this process should {\sl always} be
kinematically allowed. We demonstrate explicitly that in
our considered model its expected branching ratio can reach
$10^{-5}$ or so and lies therefore within the sensitivities
of the LEP experiments. Its existence would give rise to a
monoenergetic photon emitted with energy equal to half the
$Z$ mass. In addition we expect also to have a process
involving double majoron emission
$$
Z^0 \rightarrow J J \gamma
$$
A simple estimate indicates that the rate for this process
may be similar to that of the single emission process. This
process would give rise to a continuous photon spectrum that
could account for an excess of single-photon events at high energies.

Moreover, if kinematically allowed, there could be similar
processes involving the emission of CP even scalar bosons,
like the higgs bosons, or of massive pseudo-scalar bosons
charactericitc of supersymmetric models, e.g.
$Z^0 \rightarrow h \gamma \; , \quad Z^0 \rightarrow h h \gamma \; ,
\quad Z^0 \rightarrow A \gamma \;$, etc.
It is, indeed, quite conceivable that these rates can be quite
high due to the fact that, in these models, low masses for the
$h$-bosons are allowed, since their coupling to the $Z$ is
suppressed relative to that of the standard model.

\section{Model}

Here we adopt as a model for the spontaneous violation of R parity
the one proposed in ref \cite{MASI_pot3}. The model is characterized
by the basic superpotential terms
\beq
h_u u^c Q H_u + h_d d^c Q H_d + h_e e^c \ell H_d +
(h_0 H_u H_d - \mu^2 ) \Phi + h.c.
\label{P0}
\eeq
to which one adds the following terms
\beq
h_{\nu} \nu^c \ell H_u + h \Phi \nu^c S + h.c.
\label{P1}
\eeq
involving additional isosinglet superfields $({\nu^c}_i,S_i)$
carrying lepton numbers $(-1,1)$ respectively. The couplings
$h_u,h_d,h_e,h_{\nu},h$ are arbitrary matrices in generation
space, which explicitly break flavour conservation.
However, the form of the superpotential is restricted by imposing
the exact conservation of $total$ lepton number and R parity.
The addition of the new singlets to the minimal
$SU(2) \ot U(1)$ model \cite{SST} may lead to many novel
weak interaction phenomena \cite{BER,CON,CERN,wein}.
Most relevant for us here is the fact that their presence
allows both electroweak and R parity breaking to proceed
at the tree level. The spontaneous breaking of R parity
and lepton number is driven by nonzero VEVS for the
additional singlets \cite{MASI_pot3}
\bea
v_R = \VEV {\tilde{\nu}_{R\tau}} ~~~~~~~~~~~~
v_S = \VEV {\tilde{S_{\tau}}}
\eea
which are generated at the electroweak scale, whereas electroweak breaking
and fermion masses arise from
\bea
\VEV {H_u} = v_u ~~~~~
\VEV {H_d} = v_d
\eea
with $v^2 = v_u^2 + v_d^2$ fixed by the W mass.

The Majoron is given by the imaginary part of \cite{MASI_pot3}
\beq
\frac{v_L^2}{Vv^2} (v_u H_u - v_d H_d) +
              \frac{v_L}{V} \tilde{\nu_{\tau}} -
              \frac{v_R}{V} \tilde{{\nu^c}_{\tau}} +
              \frac{v_S}{V} \tilde{S_{\tau}}
\eeq
where $V = \sqrt{v_R^2 + v_S^2}$. Since the majoron
is mainly an \21 singlet it does not contribute to the
invisible $Z^0$ decay width. Moreover, one may easily satisfy
the astrophysical limit from stellar cooling \cite{Dear}
for $v_R = O(1\:TeV)$ and $v_L \lsim O(100 \:MeV)$.
As shown in ref \cite{MASI_pot3}, this hierarchy
between $v_L$ and $v_R$ may be achieved in a natural way.

In order to study the radiative $Z^0$ decays with majoron
emission we need the structure of the $chargino$ mass matrix,
given by \cite{MASI_pot3}
\beq
\begin{array}{c|cccccccc}
& e^+_j & \tilde{H^+_u} & -i \tilde{W^+}\\
\hline
e_i & h_{e ij} v_d & - h_{\nu ij} v_{Rj} & \sqrt{2} g_2 v_{Li} \\
\tilde{H^-_d} & - h_{e ij} v_{Li} & \mu & \sqrt{2} g_2 v_d\\
-i \tilde{W^-} & 0 & \sqrt{2} g_2 v_u & M_2
\end{array}
\label{chino}
\eeq
where $g_{1,2}$ are the $SU(2) \ot U(1)$ \gau couplings divided by
$\sqrt{2}$ and $M_{1,2}$ denote the supersymmetry breaking gaugino
mass parameters, related by $M_1/M_2 = \frac{5}{3} tan^2{\theta_W}$.
In many models, such as the one in ref \cite{MASI_pot3}, the effective
Higgsino mixing parameter $\mu$ may be given as $\mu = h_0 \VEV \Phi$,
where $\VEV \Phi$ is the VEV of an appropriate singlet scalar.

The form of this matrix applies to a wide class of models
and determines the masses of the physical charged leptons
as well as those of the charginos. As a result of R
parity breaking, the supersymmetric fermions will now mix
with the weak-eigenstate leptons. Moreover \eq{chino} also
specifies the couplings of the majoron and of the $Z^0$ to the
mass-eigenstate charged leptons and the charginos, which will
be relevant for calculating the $Z^0\rightarrow\gamma J$ decay
width.

\section{The process $e^+ e^- \ra Z^0 \rightarrow \gamma J$ }

Single photon events can be produced in our model via the
radiative $Z^0$ decays into a photon plus the invisible
majorons. At the one-loop level the single majoron
monophoton events arise from the diagrams shown in Fig. 1.
The general form of the amplitude for the process of Fig. 1
can be expressed as:
\begin{eqnarray}
V^{\mu\nu}=i \left(V_1 g^{\mu\nu} + V_2 p^{\mu} p^{\nu} + V_3 q^{\mu} q^{\nu}
+ V_4 p^{\mu} q^{\nu} + V_5 p^{\nu} q^{\mu}\right) - V_6 \epsilon^{\mu\nu
\alpha\beta} p_{\alpha} q_{\beta}
\end{eqnarray}
Imposing the on-shell conditions $p^2=0$, $q^2=M_J^2=0$,
$(p+q)^2=M_Z^2$ and requiring the gauge-invariance conservation
$V^{\mu\nu}_{ON}p_{\nu}=0$, this reduces to:
\begin{eqnarray}
V^{\mu\nu}_{ON}=i \left[V_1 \left(g^{\mu\nu} - {p^{\mu} q^{\nu} \over
pq} \right) + V_2 p^{\mu} p^{\nu} + V_5 p^{\nu} q^{\mu}\right]
- V_6 \epsilon^{\mu\nu\alpha\beta}p_{\alpha}q_{\beta}
\end{eqnarray}
In matrix element calculations form factors $V_2$ and $V_5$ do not
contribute after contraction with the photon polarization vector, hence
the effective form of the $Z^0 \gamma J$ vertex can be expressed as:
\begin{eqnarray}
V^{\mu\nu}_{ON}=i V_1 \left(g^{\mu\nu} - {p^{\mu} q^{\nu} \over
pq} \right)
- V_6 \epsilon^{\mu\nu\alpha\beta}p_{\alpha}q_{\beta}
\end{eqnarray}
Form factors $V_1$ and $V_6$ are obtained by the explicit
calculation of the appropriate 3-point Green function.
Let us denote the $Z^0\chi_i^-\chi_j^+$ and $J\chi_i^-\chi_j^+$
vertices as
\begin{eqnarray}
ie\gamma^{\mu}(A_{ij} P_L + B_{ij} P_R) ~~~~~~~~~~~~ (Z^0\chi_i^-\chi_j^+)
\end{eqnarray}
\begin{eqnarray}
e(C_{ij} P_L + D_{ij} P_R) ~~~~~~~~~~~~~~ (J\chi_i^-\chi_j^+)
\end{eqnarray}
respectively. These coupling matrices have been given
explicitly in ref. \cite{NPBTAU,RPLHC}. For definiteness and
simplicity, we will assume CP conservation. It is important
to notice that, in this case, these coupling matrices obey the
following symmetry properties:
\begin{eqnarray}
A_{ij} = A_{ji} ~~~~~~~~~ B_{ij} = B_{ji} ~~~~~~~~~ C_{ij} = - D_{ji}
\end{eqnarray}
Note that gauge invariance forces the $\gamma\chi_i^-\chi_i^+$
vertex to be diagonal in the $i,j$ indices and reads as
(electric charge $q_i=1$ in our case):
\begin{eqnarray}
ieq_i\gamma^{\mu} ~~~~~~~~~~~~~~~~~~~~~~~~~~~~ (\gamma\chi_i^-\chi_i^+)
\end{eqnarray}
We work in the mass eigenstate basis, therefore the
particles exchanged in the loop are the five physical
singly charged fermions present in our model (the two
charginos and the three charged leptons). The formulae
for the form factors $V_1$ and $V_6$ (calculated on-shell)
are the following:
\begin{eqnarray}
V_1& = &-{e^3\over 8\pi^2}\sum_{i,j=1}^5\left[m_i^2 m_j
(A_{ji}D_{ij} + B_{ji}C_{ij} + A_{ij}C_{ji}
+ B_{ij}D_{ji})c_0\right.\nonumber\\
&+& m_i (A_{ji}C_{ij} + B_{ji}D_{ij} + A_{ij}D_{ji} + B_{ij}C_{ji})
(2c_{24} + pq(c_0 + c_{12}))\\
&-& \left. m_j (A_{ji}D_{ij} + B_{ji}C_{ij}
+ A_{ij}C_{ji} + B_{ij}D_{ji})(2c_{24} + pq(c_{12}
+ 2c_{23}) + {1\over 2})\right]\nonumber\\
V_6 &=&-{e^3\over 8\pi^2}
\sum_{i,j=1}^5
\left[m_j (A_{ji} D_{ij} - B_{ji} C_{ij} - A_{ij}C_{ji}
+ B_{ij}D_{ji})c_{12}\right.\nonumber\\
&+&\left. m_i (A_{ji}C_{ij} - B_{ji}D_{ij} - A_{ij}D_{ji} + B_{ij}C_{ji})
(c_0 + c_{12})\right]
\end{eqnarray}
where $pq = M^2_Z/2$ and $c_0$, $c_{11}$ and $c_{12}$ are the usual
three point functions of t'Hooft and Veltman \cite{thvelt}, called
with arguments $(p, q, m_i^2, m_i^2, m_j^2)$; for example,
$c_0 \equiv c_0 (p, q, m_i^2, m_i^2, m_j^2)$.

One should note that in order to obtain non-vanishing amplitude
for the discussed process, R-parity must be broken in one of the
$Z^0\chi_i^-\chi_j^+$ or $J\chi_i^-\chi_j^+$ vertices. This is
clear, since the majoron arises from a \21 singlet sneutrino
and is therefore R-odd.

The decay width for the process $Z^0\rightarrow\gamma J$ can
be written as:
\begin{eqnarray}
\Gamma(Z^0\rightarrow\gamma J) = {1\over 96\pi M_Z} \left(4|V_1|^2 +
M_Z^4 |V_6|^2\right)
\end{eqnarray}
The differential and total cross sections for the process
$e^+e^-\rightarrow\gamma J$ have the following forms (with the $Z^0e^+e^-$
vertex defined as $ie\gamma^{\mu}(v-a\gamma_5)$):
\begin{eqnarray}
{d\sigma\over d\Omega} (e^+e^-\rightarrow\gamma J)
&=& {e^2\over 512\pi^2 M_Z^2\Gamma_Z^2} \left[ (v^2+a^2)(4|V_1|^2 +
M_Z^4 |V_6|^2)(1 +\cos^2\theta)\right.\nonumber\\
& -&\left. 16av M_Z^2 \mathrm{Re}(V_1V_6^\star)\cos\theta \right]
\end{eqnarray}
\begin{eqnarray}
\sigma (e^+e^-\rightarrow\gamma J)
= {e^2(v^2+a^2)\over 96\pi M_Z^2\Gamma_Z^2} \left(4|V_1|^2 +
M_Z^4 |V_6|^2 \right)
\end{eqnarray}

Note also that the $V_1$ form factor
is non-vanishing only if the CP is broken, i.e. $V_1\neq 0$
only if parameters in the lagrangian contain complex phases.
This follows directly from the symmetry properties of the
coupling matrices discussed above. We will proceed, in what
follows, assuming CP conservation, so that $V_1 = 0$.

In order to calculate the attainable values of the
the branching ratio for the 2-body $Z$ decay
to a single photon plus missing momentum in the
spontaneously broken R parity model we have
varied the unknown model parameters, which
include the standard supersymmetric parameters
$\mu$, $M_2$, $\tan\beta$,
as well as the effective R parity violating parameter
$h_\nu v_R$  over a reasonable range specified as
\begin{eqnarray}
&-500\ GeV \leq \mu \leq 500\ GeV & \nonumber \\
&30\ GeV \leq M_2 \leq 500\ GeV&\nonumber \\
&10\ \leq tan \beta \leq 40 & \nonumber \\
&1 \leq h_{\nu} \leq \sqrt{4 \pi}\  & \nonumber \\
&10 GeV \leq v_R \leq 100\ GeV &
\end{eqnarray}

We should stress that, in order to perform a systematic
analysis of the attainable values of this branching ratio
one needs to reject all points
that violate the constraints that follow from all existing
observations, including those from laboratory, cosmology and
astrophysics. They follow mostly from \neu physics and
from SUSY particle searches, and have been described in
earlier papers \cite{NPBTAU,RPLHC}.

We have found that the branching ratio for the 2-body $Z$ decay
to a single photon plus missing momentum in the spontaneously
broken R parity model can reach $10^{-5}$ or so and is, therefore,
within the sensitivities of the high luminosities possible at LEP.
This is illustrated in Fig. 2.

\section{Non-monochromatic single photon production at LEP}

In the spontaneously broken R parity model,
in addition to the standard model diagrams, there are new sources
contributing to non-monochromatic single photon production
coming from the decay $Z^0 \rightarrow JJ \gamma$ in which
two majorons are emitted.

The leading diagram
\footnote{Other mechanisms of generating non-
monochromatic single photon events exist in the
spontaneously broken R parity model. For example,
LSP pair production followed by the radiative
decay of one of them and the invisible decay
of the other $\chi^0 \ra J \nu$. Here we are
not considering this possibility.}
for this process is the same that
gives the single emission process, but removing the $R$
parity violating insertion $V_{R}$ by the second propagating
majoron and is illustrated in Fig. 3.
This amplitude
can lead to a sizeable $Z^0 \rightarrow JJ \gamma$ decay branching
ratio for sufficiently large values of $h_{\nu}$ and correspondingly
low $V_{R}$ values, in order to keep the \nt mass within the
experimentally allowed range. A simple estimate suggests that
$BR(Z^0 \rightarrow \gamma JJ) \sim 10^{-5}$ is allowed.

There is another class of diagrams contributing to the $JJ\gamma$
process. They are similar to a corresponding class of standard model
radiative diagrams leading to $Z^0 \rightarrow \nu \bar{\nu} \gamma$.
However, now the $Z^0 \nu \bar{\nu}$ vertex is replaced by the $hJJ$
vertex. These amplitudes may be assumed to be negligible in the region
of parameters where the $hJJ$ coupling is small.

Note that the photons produced in the $Z^0 \rightarrow JJ \gamma$
decay have a different energy distribution from the corresponding
bremsstrahlung $Z^0 \rightarrow \nu \bar{\nu} \gamma$ decays of the
standard model. Indeed, under quite general arguments one can write
the relevant amplitude as being proportional to a factor
$$
V_{\mu \nu}= i (V_{1} g_{\mu \nu} + V_{2} p_{\mu}p_{\nu} + V_{3} p_{\mu}
q_{1 \nu} + V_{4} p_{\mu} q_{2 \nu}
$$
$$
+ V_{5} q_{1\mu} p_{\nu} + V_{6} q_{2\mu} p_{\nu} + V_{z} q_{1\mu} q_{2\nu} +
V_{8} q_{1\mu} q_{2\nu}
$$
$$
+ \epsilon^{\mu \nu \alpha \beta}(V_{9} p_{\alpha} q_{1 \beta} + V_{10}
p_{\alpha} q_{2\beta} + V_{11} q_{1 \alpha} q_{2 \beta})
$$

Now imposing electromagnetic gauge invariance, $CP$ invariance and bose
symmetry for the two emitted majorons, one can express the on-shell amplitude
in terms of a single form factor $V_{0}$ as
\begin{equation}
A (Z^0 \rightarrow JJ \gamma ) = \epsilon_{\mu} (Q) \epsilon_{\nu} (p) V_{0}
[p^{\mu} (q_{1} + q_{2})^{\nu} - p (q_{1} + q_{2})g^{\mu \nu} ]
\label{amplitude}
\end{equation}
The $V_{0}$ form-factor is, in general, a function of the kinematical
variables which has to be determined by the calculation of
the loop diagram in Fig. 3.

As the simplest illustration we neglect this dependence and derive the
$\gamma$-spectrum following from eq.~\ref{amplitude} in the approximation of
constant
$V_{0}$. We obtain
$$
\frac{d \Gamma}{d E_{\gamma}} = \frac{|V_{0}|^{2} m_Z}{96 \pi^{3}}
E_{\gamma}^{3}
$$
\noindent
with $0 \leq E_{\gamma} \leq \frac{m_Z}{2} \; . $

Note that in this idealized limit we obtain a spectrum characterized by a
single free parameter $|V_{0}|$. In general, however, we expect that the
shape of the $\gamma$ spectrum in this case will depend upon the unknown
supersymmetric parameters that characterize the couplings in the
loop diagram in Fig. 3 in a rather complicated way.

Superimposed upon the continuous spectrum we have, in addition,
a spike at its endpoint, corresponding to the emission of the
monochromatic $\gamma $ from the $Z^0 \rightarrow \gamma J $ decay.

\section{Discussion}

Unlike the standard model, where missing energy is always
carried by a neutrino-antineutrino pair, in the spontaneously
broken R parity models, due to the existence of the light
pseudoscalar majoron, one can generate {\sl mono-energetic} photons
plus missing momentum in $Z^0$ decays. This feature is also characteristic of
other singlet majoron models with the spontaneous violation of
lepton number ocurring at a scale similar to the weak interaction
scale. The single production rates at LEP will, of course, be
model dependent. The existence of the majoron in these models is
also very interesting from the point of view of opening new
decay channels for the higgs boson which may even be
dominant with respect to the standard $b\bar{b}$ decay
mode \cite{HJJ_JoshipuraValle92}. The role of these
invisible higgs decays in the strategies to search for
the higgs bosons has been considered both for $\epm $
\cite{alfonso} as well as hadron colliders \cite{granada1}.
Many other phenomenological features of these models
have been considered \cite{granada}.

As we have seen, these models also provide
new mechanisms for generating non-monochromatic single
photons at LEP, coming from the decay $Z^0 \rightarrow JJ \gamma$
in which two majorons are emitted. These decays will lead to
a continuous photon spectrum that might account for a possible
excess of high energy photons over the expectations from
initial state radiation.

In short, we find our results encouraging from the point of view
of the recent hints from the OPAL experiment. They suggest that a
more thorough experimental investigation of single-photon events
would be welcome since it might lead, with good luck, to a way
to sort out the novel decays considered here from those with similar
characteristics, that happen within the standard model. A more
detailed theoretical study of the shape of the continuous
single photon spectrum would also be desirable.

\section*{Acknowledgements}
\hspace{0.3cm} This work has been supported by the
DGICYT Grant PB92-0084, by the EEC grant CHRX-CT93-0132,
as well as a postdoctoral fellowship (J.R.) and Acci\'on
Integrada HP-94-053. We thank Michael Dittmar for very
helpful discussions.

\newpage
{\Large \bf Figure Captions}
\noi

Fig. 1: \\
Feynman diagrams contributing to the monochromatic
photon emission process $Z^0 \rightarrow \gamma J$.

Fig. 2: \\
Allowed values for the branching ratios for the process
$Z^0 \rightarrow \gamma J$, as a function of $m_{\nu_{\tau}}$,
for various values of $tan \beta$ and $v_R$. Other supersymmetric
parameters were randomly chosen as described in text.

Fig. 3: \\
Higher order diagram contributing to the non-monochromatic
single photon production.

\end{document}